\documentclass[usenatbib,usegraphicx]{mn2e}
\usepackage{times}
\usepackage{graphicx, epsfig}
\usepackage[fleqn]{amsmath}
\usepackage{amsfonts}
\usepackage{amssymb}

\title[IFS of Centaurus A]{Integral-field spectroscopy of Centaurus A
  nucleus} \author[Davor Krajnovi\'c et al.]  {Davor
  Krajnovi\'c,$^1$\thanks{E-mail: dxk@astro.ox.ac.uk} Rob Sharp,$^2$
  Niranjan Thatte,$^1$\\
$^1$Denys Wilkinson Building, University of Oxford, Keble Road, OX1 3RH, UK \\
$^2$ Anglo-Australian Observatory, Epping NSW 1710, Australia 
}

\date{Accepted 2006 September 19. Received 2006 September 19; in original form
2006 July 21}
\pagerange{\pageref{firstpage}--\pageref{lastpage}}
\pubyear{2006}

\newcommand{\kms}{\>{\rm km}\,{\rm s}^{-1}}

\newcommand{\ud}{\mathrm{d}}
\def\cirpass{\textsc{cirpass}}
\def\idl{\textsc{idl}}
\def\iraf{\textsc{iraf}}
\def\J{\emph{J}}
\def\H{\emph{H}}
\def\ngc5128{NGC\,5128}

\def\aj{AJ}             
\def\araa{ARA\&A}       
\def\apj{ApJ}           
\def\apjl{ApJ}          
\def\aap{A\&A}          
\def\aapr{A\&A~Rev.}    
\def\mnras{MNRAS}       
\def\pasp{PASP}         

\begin{document}
\label{firstpage}
\maketitle

\begin{abstract}
  We report integral-field spectroscopic observations with the
  Cambridge Infra-Red Panoramic Survey Spectrograph (CIRPASS) mounted
  on the GEMINI South telescope of the nucleus of the nearby galaxy
  NGC~5128 (Centaurus A). We detect two-dimensional distributions of
  the following emission-lines: [PII], [FeII] and Paschen $\beta$. We
  compare our observations with previously published radio
  observations (VLA) and archival space-based near-infrared imaging
  (HST/NICMOS) and find similar features, as well as a region of high
  continuum coinciding with the jet (and its N1 knot) at about
  2\arcsec\, North-East of the nucleus, possibly related to
  jet-induced star formation.  We use the [FeII]/[PII] ratio to probe
  the ionisation mechanism, which suggests that with increasing radius
  shocks play an increasingly important role. We extract spatially
  resolved 2D kinematics of Pa$\beta$ and [FeII] emission-lines.  All
  emission-line regions are part of the same kinematic structure which
  shows a twist in the zero-velocity curve beyond $\sim 1\arcsec$ (for
  both Pa$\beta$ and [FeII]). The kinematics of the two emission-lines
  are similar, but the Pa$\beta$ velocity gradient is steeper in the
  centre while the velocity dispersion is low everywhere. The velocity
  dispersion of the [FeII] emission is relatively high featuring a
  plateau, approximately oriented in the same way as the central part
  of the warped disk. We use 2D kinematic information to test the
  hypothesis that the ionised gas is distributed in a circularly
  rotating disk. Assuming simple disk geometry we estimate the mass of
  the central black hole using Pa$\beta$ kinematics, which is
  consistent with being distributed in a circularly rotating disk. We
  obtain M$_\bullet = 8.25^{+2.25}_{-4.25} \times 10^{7}$ M$_{\odot}$,
  for $PA = -3\degr$ and $i=25\degr$, excluding the M$_\bullet -
  \sigma$ relation prediction at a $3\sigma$ confidence level, which
  is in good agreement with previous studies.
\end{abstract}


%
%

\section{Introduction}
\label{s:intro}

NGC 5128 is the king galaxy of the southern hemisphere, if sovereignty
is judged by the number of studies devoted. It is no wonder that this
is the case since Centaurus A, or Cen A, as it is commonly referred
to, is the fifth brightest galaxy in the sky
\citep[RC2,][]{1976rcbg.book.....D} usually classified as an E0p or
S0p.  It hosts a powerful X-ray and radio jets
\citep{2003ApJ...593..169H, 2004ApJ...612..786E} and a low-luminosity
AGN \citep{1996ApJ...466L..63J, 2001MNRAS.324L..33C}. Cen A suffered a
recent major merger that shaped its present appearance, leaving shells
at large radii \citep{1983ApJ...272L...5M} and a conspicuous polar
dust lane in the centre \citep{1979AJ.....84..284D}, consistent with
being a thin warped disk
\citep{1992ApJ...391..121Q,1992ApJ...387..503N,1993ApJ...412..550Q,
  QU06}. It is, however, the proximity that makes this object truly
special.  Its most recent distance estimate places it at $3.42 \pm
0.25$ Mpc \citep{Fe06}, while other recent studies give: $3.40 \pm
0.15$ Mpc \citep{1998A&ARv...8..237I}, $4.2 \pm 0.3$Mpc
\citep{2001ApJ...546..681T} and $3.84 \pm 0.35$ Mpc
\citep{2005ApJ...631..262R}. Being so close, Cen A is also the closest
AGN and recent merger, a great case for testing our understanding of
processes that shape galaxy formation and evolution \citep[for
  detailed information about the galaxy see the review
  by][]{1998A&ARv...8..237I}.

The thick dust lane that crosses Cen A, making it a show case for
astronomical PR, hides the centre of the galaxy and its
constituents. Within the AGN paradigm \citep{1993ARA&A..31..473A}, it
is expected that there is a suppermassive black hole at the bottom of
Cen A's potential well. Indeed, recent studies, 'peering through the
dust' with near-infrared spectrographs, investigated this assumption
and found that it is necessary to place a dark object in the nucleus
with a mass of: M$_\bullet = (2^{+3.0}_{-1.4}) \times 10^8$M$_{\odot}$
for inclination $i> 15\degr$ \citep[ground-based gas
  kinematics,][]{2001ApJ...549..915M}, M$_\bullet =(2.4^{+0.3}_{-0.2})
\times 10^8$M$_{\odot}$ for an edge-on model, M$_\bullet
=(1.8^{+0.4}_{-0.4}) \times 10^8$M$_{\odot}$ for $i=45\degr$ and
M$_\bullet =(1.5^{+0.3}_{-0.2}) \times 10^8$M$_{\odot}$ for
$i=20\degr$ \citep[ground-based stellar
  kinematics,][]{2005AJ....130..406S}, M$_\bullet =(6.1^{+0.6}_{-0.8})
\times 10^7$M$_{\odot}$ at $i=45\degr$ \citep[ground-based adaptive
  optic assisted observations of gas
  kinematics,][]{2006ApJ...643..226H} and M$_\bullet =(1.1 \pm 0.1)
\times 10^8$M$_{\odot}$ for $i=25\degr$ and M$_\bullet
=(5.5^{+0.7}_{-0.6}) \times 10^7$M$_{\odot}$ for $i=45\degr$
\citep[ground- and space-based observations of gas
  kinematics,][]{2006A&A...448..921M}.  The last study also
constrained the size of the central massive object to r $\sim0.6$ pc
suggesting that it is indeed a black hole. The importance of a robust
measurement of Cen A's M$_\bullet$ is highlighted in the fact that the
above masses place the black hole in Cen A approximately between 2
times and an order of magnitude above the M$_\bullet -\sigma$ relation
\citep{2000ApJ...539L..13G, 2000ApJ...539L...9F,
  2002ApJ...574..740T,2005SSRv..116..523F}. Although the large scatter
in the relation and the low number statistics might be misleading, it
is also possible that Cen A, as a recent merger, is a special case
\citep{2005AJ....130..406S}.

Each of these studies brought an improvement in spatial and spectral
resolution, but they were all based on observations with long-slits
along a few position angles. For construction of dynamical models,
whether stellar or gaseous, it is crucial to have sufficient
information about the behaviour of the constraints. In the case of
stellar dynamical models, it is necessary to observe the kinematics at
all positions (within $\sim 1 R_e$) to robustly constrain the orbital
distribution \citep{2005CQGra..22S.347C, 2005MNRAS.357.1113K, SH06},
while for gaseous models the geometry of the gas disk can only be
confirmed by two-dimensional mapping and only a well-ordered gas disk
can be used to reliably determine the enclosed mass
\citep{2002ApJ...567..237S}. 

The nucleus of Cen A is active and the dusty and gaseous polar disk is
significantly warped from small to large scales, where the disk
alternates between having the southern side (within 60\arcsec) and
northern side (beyond 100\arcsec) nearest to the observer (see Fig.~11
of \cite{1992ApJ...387..503N} and Fig.~5 of \cite{QU06}). When
constructing dynamical models it is important to understand in what
way these two properties shape the observations. The ionisation of
emission-lines could be under a strong influence of the AGN, while the
kinematic signature of a disk warp could be hard to recognise from
long-slit measurements only.

In this paper we present the first observations with an integral-field
spectrograph of the nucleus of Cen A in the near-infrared. In
Section~\ref{s:obs} we describe the observations and the data
reduction. In Section~\ref{s:res} we analyse the data and compare them
with existing observations. We describe the kinematical properties of
the nuclear disk-like structure in Section~\ref{s:kinem}. In
Section~\ref{s:dyn} we discuss the evidences for the central black
hole. Section~\ref{s:con} summarises our conclusions.

Throughout the study we assumed a distance to Cen A of 3.5 Mpc to be
consistent with the dynamical studies mentioned above.

%
%

\section{Observations and data reduction}
\label{s:obs}

We used the Cambridge Infra-Red Panoramic Survey Spectrograph,
\cirpass\, \citep{2004SPIE.5492.1135P}, mounted at the Gemini South
telescope, during the March 2003 queue scheduled observing
campaign. Some general observational details are given in
Table~\ref{Observations}.

\subsection{CIRPASS data reduction}
\label{ss:reduc}

The \cirpass\ spectrograph, housed in an insulated cold room and
cooled to -42$^\circ$C, is a fibre fed instrument with observing
capabilities in the \J\ and \H\ bands. Operating at a moderate
resolution (R$\sim$3,000), the night sky OH emission spectrum is
sufficiently dispersed to allow one to capitalise on its intrinsic
darkness between emission lines.

\cirpass\ has both integral field \citep[e.g.][]{2004AN....325..108S}
and multi-object observing modes \citep[e.g.][]{2004MNRAS.354L...7D,
  2006MNRAS.tmp..641D}.  The IFU consists of a 490 element
macrolens\footnote{\cirpass\ employs a macro lens array rather than a
  micro lens array, to avoid Focal Ratio Degradation (FRD) losses
  \citep{1998SPIE.3354..668E, 1998SPIE.3355..926K}} array arranged in
a broadly rectangular structure, with hexagonal lens packing.  Through
the use of re-imaging optics, in the focal plane, two lens scales
where available at Gemini, 0\farcs36 and 0\farcs25, giving fields of
view of 13$\times$4\farcs7 and 9\farcs3$\times$3\farcs5 respectively.
We have used the 0\farcs25 scale to study the central region of Cen A
with the spectral resolution of $\sim2.2$ \AA/pixel (FWHM) and
resolution element of 1.8 pixels, in the spectral range between $1.1 -
1.35~\mu m$.

Data is recorded as series of on- and off-target exposures (each 900
s). Off-target sky exposures were taken with a sufficient offset
(600\arcsec) from the Cen A nebulosity to avoid contamination from the
galaxy. Data reduction was performed following the recipe outlined in
the \cirpass\ data reduction cookbook\footnote{The \cirpass\, data
  reduction software and documentation is available from {\tt
    http://www.ast.cam.ac.uk/$\sim$optics/cirpass/index.html}}.  The
\cirpass\ \iraf\ data reduction tasks where used along side a suite of
custom software written in the \idl\ programing environment.

Spectra are extracted form the beam-switched science frames, the sky
frame and dome flats taken immediately prior to or after the science
frames.  Standard star frames are also extracted, with a beam switch
pair placing the star at either end of the IFU to allow sky
subtraction without the need for offset sky frames.  An optimal
extraction routine is used to allow for the crosstalk between fibres.

Extracted spectra are then flat-fielded, using the normalised
extracted flat field image.  This step cannot be performed before the
extraction as it would remove much of the information on the fibre
profiles which we wish to extract.  A wavelength solution and 2D
transformation to a common wavelength scale is performed using the
\iraf\ tasks \emph{identify, reidentify, fitcoo} and \emph{transform}.
An average OH night sky spectrum is used for the wavelength solution,
using only stronger lines from the line list of
\cite{1993PASP..105..940M} and free from blends. The average flat
field response is then multiplied back into the data to suppress the
spectrum of the flat field lamp.

A relative flux calibration is derived using observations of the
standard star Hip066842. For the star, a representative black-body
spectrum is assumed and used to generate a spectral response function
and telluric absorption correction, after interpolating across strong
stellar absorption features. The spectrum of Vega from the {\tt IRAF
  SYNPHOT} packages is used as a reference for the magnitude scale.
Both the observed and the reference Vega spectrum are integrated over
a clean region (away from telluric and strong stellar features) of the
observed spectral range to find the calibration zero-point.

\begin{table}
\caption{\label{Observations}Observational details}
\begin{tabular}{cccc}
\hline
\textbf{Date} & \textbf{Band} & \textbf{Exp. time } & \textbf{Seeing}\\
\hline
13. 03. 2003.& \J & 3$\times$900 s & ~0\farcs6\\
19. 03. 2003.& \J & 6$\times$900 s & ~0\farcs5\\
\hline
\end{tabular}
\newline
Notes : We adopt the date at the start of the Chilean night,
rather than the Gemini default which would move all dates one day
later.
\end{table}

\subsection[]{Mosaicing and alignment}
\label{ss:mosaic}

During typical \cirpass\ observations, dither offsets on target are
performed using integer lens offsets to allow straight-forward
combination of the observational data sets. However, data in each
passband was taken using a number of different offset strategies.
Alignment and combination is complicated further by the hexagonal
lenslet array.  In order to align and stack the various data sets we
therefore resample the data onto a regular Cartesian grid, determine
optimal image alignments via cross-correlation, and combine the data
using integer pixel shifts.  The process is as follows.

For data taken on different nights the wavelength solution was
significantly different. We therefore first correct all data in a
specific band to a common wavelength solution. We achieve this using
an \idl\ implementation for the STARLINK FIGARO data reduction task
SCRUNCH which allows accurate variance array propagation on
rebinning.

A Cartesian grid is generated which completely encloses the individual
\cirpass\ IFU frames.  Since the entire Cen A data set was recorded
using a position angle ($PA$) of zero, no account need be taken of
rotation of the IFU coordinate frame.  The pitch of the regular grid
is chosen to oversample the IFU lens scale. Oversampling by a factor
of three was found to give an excellent compromise between preserving
the data integrity while keeping the output data set (which has of the
order of 1000 wavelength channels) to a tractable size.  A spectrum is
assigned to each element of the regular array based on proximity to
the elements of the IFU array.

A set of integrated images is generated from these data cubes by
summing along the dispersion axis, avoiding wavelength ranges subject
to questionable data due to spectral curvature.  These summed images
are then cross-correlated, using elements of the \idl\ \emph{astrolib}
libraries \citep{1993adass...2..246L}.  Initial shift estimates are
generated by eye from the summed images.  The generated shifts are
then used to apply integer pixel shifts to the resampled data cubes.

Variation in the recorded intensity of spectra within the individual
observations was at the 10\% level during each night and higher
between nights (due primarily to the difficulty associated with
observing accurate standard stars, over a range of airmasses, before,
during and after each data set is recorded).  We therefore derive a
multiplicative scaling factor to apply to each data set before the
data is combined.  The scaling factors are derived through measurement
of the mean quotient between each data set and a reference data set,
for all spectra which are found in the complete overlap region between
all IFU pointings ($\sim$33\%\ of the data).

After scaling each data cube, the data are mean combined using a
3-sigma clipping algorithm to reject cosmic rays, which are still
present to some extent in the data at this stage (as there are too few
degenerate frames for a simple sigma clipping of the unextracted data)
both as positive and negative features in the data due to the
beam-switched sky subtraction. The resulting pixels scale is a factor
of 3 smaller than the actual spatial resolution. We return to the
original \cirpass\, pixel size of 0\farcs252 by averaging spectra
within neighbouring $3\times3$ pixels. The error arrays were
propagated by averaging in quadrature.

\subsection{Binning}
\label{ss:binning}

A minimum signal-to-noise ratio ({\it S/N}) is required to reliably
measure the properties of spectral lines. To achieve this, we
spatially binned our merged data using an adaptive algorithm of
\citet{2003MNRAS.342..345C} creating compact and non-overlapping bins
with uniform {\it S/N} in faint regions. A desirable value of {\it
  S/N} depends on the purpose of the analysis. Similarly, a way to
estimate the {\it S/N} can be different. Usually, if each pixel in a
spectrum has a signal $S_i$ and a measurement uncertainty (noise)
$N_i$, then the {\it S/N} of the spectrum of $n$ pixels is given by

\begin{equation}
\label{eq:sn}
(S/N)=\frac{\frac{1}{n}\sum_i^n S_i}{\sqrt{\frac{1}{n}\sum_i^n N^2_i}}.
\end{equation}

\noindent In this approach the main contribution to the signal comes
from the continuum (stellar) light and it is used for measurement of
stellar kinematics.

On the other hand, emission-lines do not necessarily follow the level
of the stellar continuum and when estimating their {\it S/N} it might
be more useful to estimate the 'signal' of the emission-line itself
compared to the local noise of the spectral region surrounding the
line. In this way, one is not including the information on the
continuum and that is reflected in the resulting bins, which then
directly trace the emission-lines. Following \citet{2005FathiThesis}
and \citet{2006MNRAS.366.1151S}, we define amplitude-to-noise ratio
({\it A/N}) for each emission-line (continuum subtracted spectra),
where the amplitude ($A$) of an emission line is estimated as a median
of the three largest pixel values in the emission-line, and the noise
($N$) is defined as the standard deviation of the spectra in the
spectral region not contaminated by the presence of emission-lines.

Since emission-lines are not necessarily present everywhere over the
field of view, we used both binning approaches. The {\it S/N} based
binning is used to present the data, analyse the global properties and
compare them with other observations (Section~\ref{s:res}).  We used a
target {\it S/N} of 20. The {\it A/N} based binning is used to measure
the kinematics of individual lines (Section~\ref{s:kinem}).  We
experimented with different ratios and found, similarly to
\citep{2006MNRAS.366.1151S}, that {\it A/N} = 2.5 is the most
appropriate to detect lines and reduce the loss of spatial
resolution. Both {\it S/N} and {\it A/N} were chosen such as to
minimise the degradation of spatial resolution in the central $\sim
1\arcsec$.

\subsection{Additional data}
\label{ss:addata}

In addition to {\tt CIRPASS} data, we also used other data sets to
supplement our analysis. These include ground-based optical and radio
observations and space-based optical and X-ray imaging. We used
reduced images of Cen A obtained by the Very Large Array (VLA) in
A-array configuration and with the Chandra X-ray Observatory, kindly
provided by M. Hardcastle.  Both images were observed, reduced and
presented in \citet{2003ApJ...593..169H}. We also used archival
near-infrared observations of the central region of Cen A with the
Hubble Space Telescope (HST). These include broad and narrow band
imaging using NICMOS camera 2 (0\farcs075 pixel$^{-1}$) with F187N
(Paschen $\alpha$) and F222W (K-band) filters. These images were
previously presented in \citet{1998ApJ...499L.143S} and
\citet{2000ApJ...528..276M}.  Finally, in order to constrain the mass
at a larger radius we used a 2MASS K-band image of the galaxy.

\begin{figure}
  \includegraphics[width=\columnwidth]{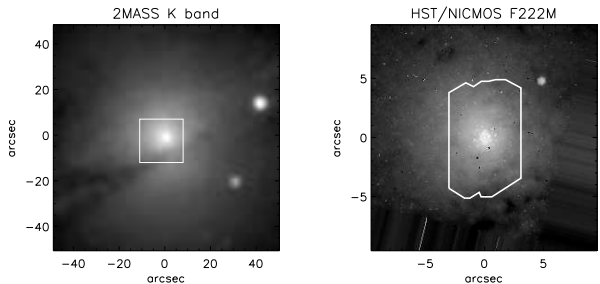}
        \caption{\label{f:foot} Broad band observations of central
          region of Cen A. Left: ground-based 2MASS K-band image; the
          over-plotted square shows the size and orientation of the
          right image.  Right: HST/NICMOS F222M image with
          over-plotted footprint of the \cirpass\, field-of-view.  On
          both images North is up and East to the left. HST image has
          been rotated accordingly. }
\end{figure}

%
%

\section{Results}
\label{s:res}

Our observations cover the nucleus of Cen A. The orientation and the
size of the merged \cirpass\, field can be seen in
Fig.~\ref{f:foot}. In this section we present the \cirpass\, data cube
and compare it with other observations.

\subsection{Nuclear spectrum}
\label{ss:nuc}

\begin{figure}
  \includegraphics[width=\columnwidth]{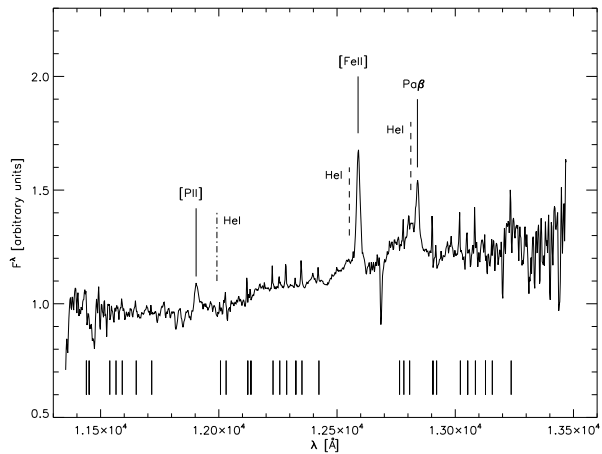}
        \caption{\label{f:spec} Total spectrum from a circular
          aperture of $0\farcs5$ radius, centred on the nucleus. Three
          emission-lines used in this study are labelled at the
          wavelength redshifted for 543 $\kms$. The vertical dashed
          lines show the position of two candidate lines contributing
          to the blue wings of detected emission-lines. There is no
          evidence for existence of HeI$\lambda11970$ line (dot-dashed
          line). The vertical lines at the bottom of the spectrum show
          positions of the sky lines in the observed wavelength
          region.  }
\end{figure}

All spectra within a circular aperture of $0\farcs5$ radius were
combined to form the nuclear spectrum presented on
Fig.~\ref{f:spec}. There one can see the three major emission-lines
which are of interest for this study.  They are, in order of their
observed relative strengths: [FeII] $\lambda$12567, Paschen $\beta$
$\lambda$12818 and [PII] $\lambda$11882 \citep{1990A&A...240..453O}.
The observed lines are redshifted, with respect to their laboratory
values, by 22 \AA, corresponding to a velocity of 543 $\kms$. The
first two lines were already observed by \citet{2001ApJ...549..915M}
having the ratio of $\sim1.7$, but the observation of the [PII] line
is, to our knowledge, the second extragalactic measurement of this
emission-line, the first being in NGC~1068 by
\citet{2001A&A...369L...5O}. The blue wings of the [FeII] and
Pa$\beta$ emission-lines could be attributed to HeI $\lambda$12530 and
HeI $\lambda$12790 emission. There is, however, no evidence for
presence of HeI$\lambda11970$ emission, which suggests HeI is
generally not strong. The origin of the red wing of the [PII] is not
known. The detected emission-lines are not contaminated by sky lines,
except a minor contribution from the sky lines to the bluest part of
the Pa$\beta$ wing.

The relative strengths of the lines can be used to determine the
nature of the ionisation mechanism. We estimated the strength of the
emission by fitting double Gaussians (see Section~\ref{ss:dgaus} for
more details) within apertures of increasing diameter, centred on the
nucleus. Table~\ref{t:ratio} shows the measurement of the ratio of the
lines. Within a 0\farcs25 aperture [FeII]/Pa$\beta \sim 2.3$, while
[FeII]/[PII] $\sim 3.0$. As the aperture increases the
[FeII]/Pa$\beta$ ratio fluctuates between 2.1 and 2.8, being typical
of active galaxies or supernova remnants, as can be seen from
observations and line diagnostic diagrams
\citep[e.g.][]{1996MNRAS.283..777S,1997ApJ...482..747A,
  2000ApJ...528..186M, 2004A&A...425..457R}. The active nucleus in Cen
A is an obvious source of [FeII] emission-lines, and although it is
possible that a fraction of the [FeII]/Pa$\beta$ ratio is caused by
star-formation (e.g. supernova remnants in the star forming regions),
the fact that [FeII] emission is associated with shock, also points
towards a mass outflow from the nucleus (e.g. the jet) and resulting
shocks as the source of ionisation \citep{1988ApJ...328L..41K}.

The [FeII]/[PII] ratio, however, shows a clearly increasing trend,
reaching 4.7 at the largest aperture. This can be used to asses the
importance of shocks in the total ionisation mechanism of [FeII]
lines. \citet{2001A&A...369L...5O} found that the [FeII]/[PII] ratio
is a good indicator for distinguishing the origin of the
emission-lines, especially if they are ionised through
photoionisation by soft X-rays or shocks propagating through the
interstellar medium. The authors state that in regions with
[FeII]/[PII] $\lesssim 2$ gas is photoionised, while if [FeII]/[PII]
$\gtrsim 20$ the material is excited by collisions in fast shock
waves. The observed values of the ratio are between these two limiting
values, although closer to the upper limit for photoionisation,
suggesting the AGN X-rays emission as the source of
photoionisation. Still, the rising values argue that shocks play an
increasingly important role in ionising the interstellar medium
further away from the nucleus. This could be reflected in the motion
of gas clouds, where a more settled motion should be observed in the
central regions.

\begin{figure}
  \includegraphics[width=\columnwidth]{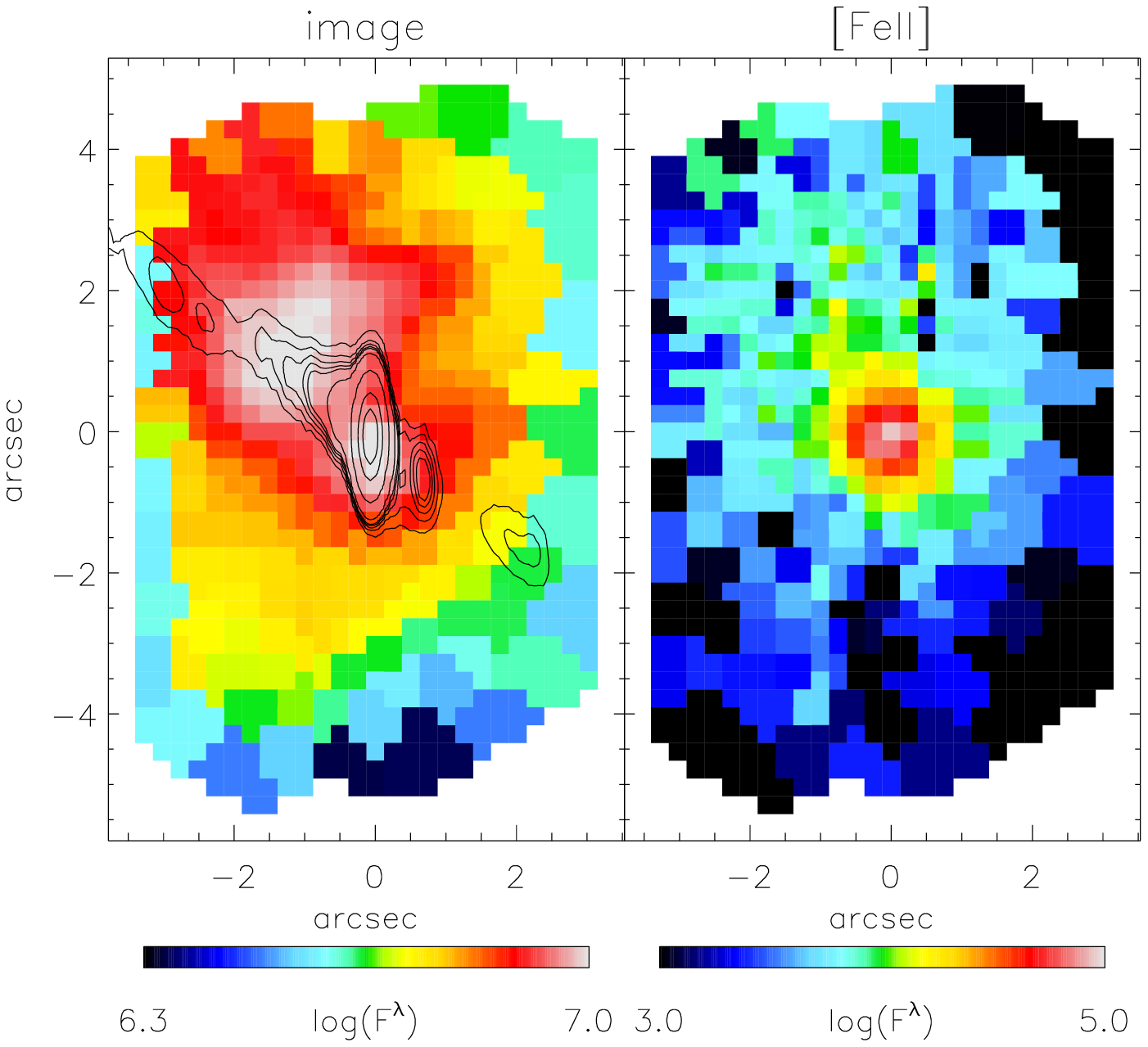}
  \includegraphics[width=\columnwidth]{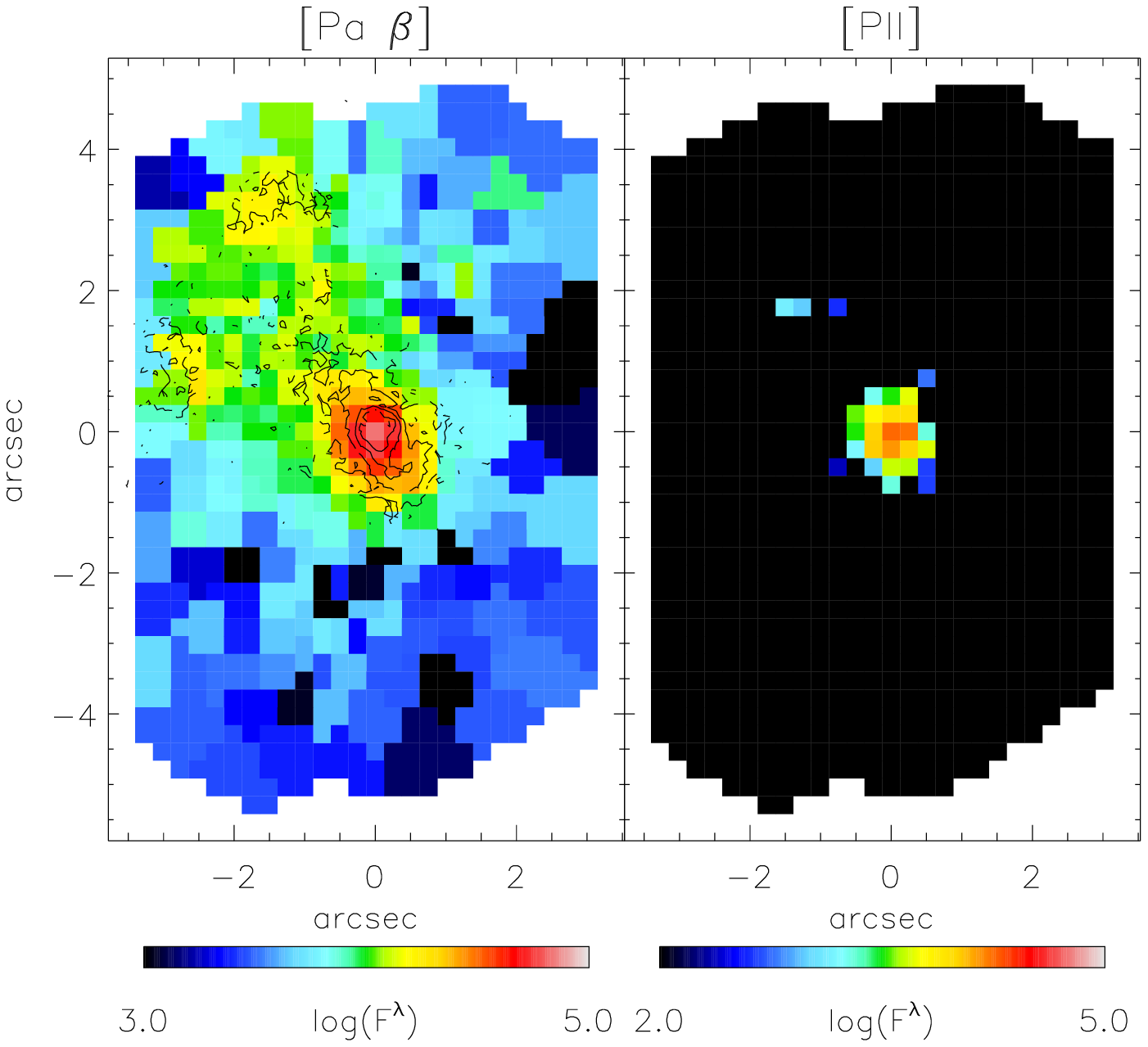}
       \caption{\label{f:flux2D} Maps obtained by summing the data
         cube along the spectral dimension after binning to $S/N =
         20$. The maps are oriented such that North is up and East to
         the left. From left to right: summed spectra with
         over-plotted VLA radio observations, spectra summed between
         12554 - 12622 \AA\, ([FeII] line), spectra summed between
         12807 - 12874 \AA\, (Pa $\beta$ line) with over-plotted
         Pa$\alpha$ emission from archival HST/NICOMS observations,
         spectra summed between 11870 - 11937 \AA\, ([PII] line). Note
         the change in the scale for the [PII] map, the agreement
         between Pa$\alpha$ and Pa$\beta$ structures and the
         coincidence of the N1 knot in the jet with the North-East
         flux peak in our observations. Flux $F^\lambda$ is in units
         of $10^{-17}$ ergs/s/m$^2$/\AA.}
\end{figure}

\begin{table}
   \caption{Emission-line ratios within circular apertures of
   different sizes centred on the nucleus}
   \label{t:ratio}
$$
   \begin{array}{ccc}
       \hline
       \noalign{\smallskip}
       $aperture$ & $[FeII]/Pa$\beta  &$[FeII]/[PII]$\\
      (1)&(2)&(3) \\
       \noalign{\smallskip}
       \hline
        0.25 & 2.3 & 3.0 \\
        0.5  & 2.8 & 3.5 \\
        0.75 & 2.1 & 3.7 \\
        1.0  & 2.2 & 4.2 \\
        1.25 & 2.6 & 4.7 \\
       \noalign{\smallskip}
       \hline
   \end{array}
$$ {Notes -- Col.(1): Radius of aperture [arcsec]; Col.(2): ratio of
     [FeII]/Pa$\beta$ flux Col.(3): ratio of [FeII]/[PII] flux.}
\end{table}

\subsection{Extent of the emission}
\label{ss:extnt}

Figure~\ref{f:flux2D} presents the spatial extent of the J-band
CIRPASS observations of Cen A. We binned the final merged data cube to
an {\it S/N} of 20. The binned spectra were continuum subtracted after
fitting 4th order polynomials to emission-free parts of spectra. The
emission-lines presented in Fig.~\ref{f:flux2D} were measured by
summing the continuum subtracted spectra along the wavelength
dimension in limited spectral regions centred on the lines themselves
(see the caption of Fig.~\ref{f:flux2D} caption for details).

Maps show two spatially separated regions of high integrated flux, one
in the nucleus (for the determination of the central pixel see
Section~\ref{ss:comp}) and the other in the North-East region (NE),
about 2\arcsec away from the nucleus. The detected emission-lines are
mostly confined to the central $1\arcsec$, but Pa$\beta$ and [FeII]
emission also extend towards the NE.  Pa$\beta$ is also present in two
blobs about 3\arcsec away from the nucleus. The weakest emission comes
from [PII]. This emission-line falls off rapidly and outside
1\arcsec\, the summed spectra are dominated by noise, but one can
obtain information on somewhat larger radius (e.g. within 1\farcs25)
by fitting the spectral line as done in Section~\ref{ss:dgaus}.

\subsection{Comparison with observations on different wavelengths}
\label{ss:comp}

In order to determine the position of the nucleus in the CIRPASS data,
we compared our observations with HST imaging.
\citet{1998ApJ...499L.143S} identified the nucleus with the peak in
Pa$\alpha$ emission, which also corresponds with the peak visible on K
band (F222W) NICMOS images and the position of the radio nucleus. We
find the position of Cen A's nucleus in our observations by assuming
that the Pa$\alpha$ and Pa$\beta$ emission have the same spatial
origin and by cross-correlating the NICMOS F187N narrow band image
with an image constructed from our unbinned data cube summed over the
spectral region of the Pa$\beta$ emission.

The spatial distribution of the two Paschen lines matches well, as it
can be seen in Fig.~\ref{f:flux2D}, where the NICMOS Pa$\alpha$ image
is over-plotted on the binned Pa$\beta$
image. \citet{1998ApJ...499L.143S} determined the position angle of
the elongation to be $\sim33\degr$. Also, the two off centred blobs
are well matched in this comparison. The extended distribution of
Pa$\alpha$ emission is also traced by Pa$\beta$ emission.

We also compared the CIRPASS observation of Cen A with existing radio
and X-ray observations \citep{2003ApJ...593..169H}. Most of the action
in the radio and X-ray happens at larger radii. This is especially the
case for X-rays where the central region is featureless, and the very
centre is saturated. The existence of X-rays, however, suggests a
possible excitation mechanism for the observed spectral lines.

Radio observations trace a jet with a position angle of 51\degr
\citep{1992ApJ...395..444C} down to the radio nucleus. Radio contours
are over-plotted on a CIRPASS image obtained by summing all the
spectra along the wavelength dimensions (Fig.~\ref{f:flux2D}). The
direction of the radio jet coincides in projection with the NE region
of high flux. The jet crosses the region of high flux loosing its
energy: at the start of the region the energy in the jet is 10
mJy/beam and at the end the emitted energy is about 1 mJy/beam (the
lowest plotted contour). This drop in jet's intensity was used to
define knot N1 by \citet{1983ApJ...273..128B}. The direction of the
jet is about 18\degr away from the orientation of the
\citet{1998ApJ...499L.143S} Pa$\alpha$ disk, but between the two
Pa$\beta$ (and Pa$\alpha$) blobs, touching the top of the East blob at
the position of the kink in the jet. The kink in the jet could be
explained as a deflection of the jet from a cloud complex which is now
glowing in Pa$\beta$ light.

\begin{figure}
  \includegraphics[width=\columnwidth]{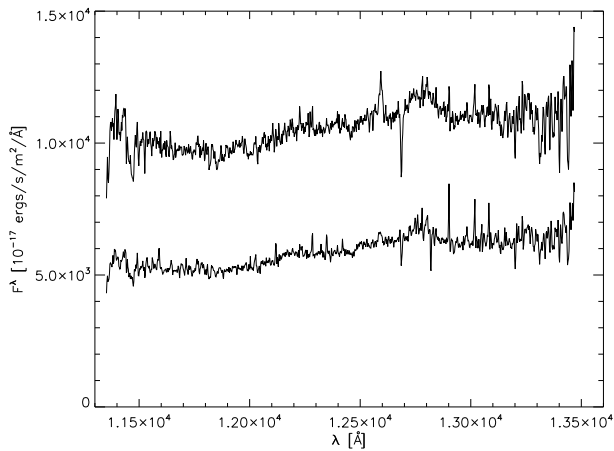}
        \caption{\label{f:2spec} An example of two spectra, one from
          the North-East region of high continuum level (top spectrum)
          and the other from the opposite South-West region with
          typical a continuum level (bottom spectrum). The spatial
          coordinates are ($-1\arcsec$, $-1\farcs26$) and ($1\farcs1$,
          $-1\farcs7$), for top and bottom spectrum, respectively.}
\end{figure}

\subsection{A hole in the extinction map or a jet-induced star formation region?}
\label{ss:sf}

The most striking feature of the reconstructed images is the secondary
peak in the integrated light which, in projection, is co-spatial with
the radio jet, bounded by the two Pa$\beta$ blobs and arises at the
end of knot N1. Cen A was the first object for which it was suggested
that star formation in certain regions was jet-induced
\citep{1981ApJ...247..813G, 1983ApJ...273..154B}, soon followed by the
Minkowski object \citep{1985ApJ...293...83V}, and numerous other
objects with different environments and jet powers at low and high
redshifts.  The connection between the suggested nuclear star
formation and the radio jet is, however, not as straight forward as at
large scales, where blobs of optical emission and enhanced continuum
correspond to specific jet structures. In this case, the end of a jet
knot is followed by an increase in the total near-infrared flux,
mostly coming from the continuum (in the same region there are no
strong emission-lines), and bounded with regions of detectable
emission: North and East Pa$\beta$ blobs and the
nucleus. Interestingly, the East Pa$\beta$ blob coincides with the
kink in the radio jet.

There are a few possible explanations for these observations. The two
most likely are that the NE flux peak is related to a jet-induced star
formation or that it coincides with the location of a hole in the
extinction. In the first scenario, star formation could happen under
the influence of the jet as it plows through the inter-stellar medium
pushing gas clouds away from its path, and resulting in a region with
higher stellar continuum. In the same time, Pa$\beta$ blobs could be
ionised from the central engine or the jet, provided that they are on
the unobscured path of non-thermal photons. In the second scenario,
the NE flux peak is visible because it is under an extinction hole. In
addition to that, the North and East Pa$\beta$ blobs may also
originate in star forming regions where young stars provide enough
photons to ionise gas clouds.

The extinction hole is supported by HST/WFPC2 F814W images
\citep[third panels on Figs.~5 and~8 of][]{2000ApJ...528..276M} on
which the nucleus (a faint point source) and the less obscured NE
region, which looks like a low extinction hole, are separated by a
dust lane. The dereddend F814W image does not have similar
features. The F222M image (K-band, presented on the same figures),
which is less influenced by obscuring material shows only a marginal
increase of flux in the NE region. The spectra from NE (high
continuum) and SW (low continuum) regions are, however, very similar
(Fig.~\ref{f:2spec}), NE spectra being systematically higher, without
any suggestion of change in the spectral slope, as it would be
expected from regions of different extinction.

Assuming that jet induced star formation is the source of the high
continuum and the Pa$\beta$ emission in the blobs, it is possible that
the North Pa$\beta$ blob was also once induced by the jet (the jet is
currently on the East Pa$\beta$ blob), which precessed down with time
(in projection) from $PA = \sim 25\degr$ to present $PA =\sim
50\degr$.

It is hard from the present observations to disentangle the true
origin of the properties of the North-East region. The precession of
the jet is perhaps a far-fetched assumption, but our data, which are
not complete enough for a decisive conclusion, support marginally more
the jet-induced star formation in the continuum region, where an
extinction hole can also contribute to the detection.

%
%

\section{Kinematical properties of the nuclear disk}
\label{s:kinem}

In the previous section we binned to a {\it S/N} ratio of 20 to
present the data and discus the main trends. In this section we
concentrate on the three observed emission-lines and use the {\it A/N}
ratio based binning. We used a value of {\it A/N} = 2.5 for all three
lines.

\subsection{Narrow and broad components}
\label{ss:dgaus}

The shape of the emission-lines on Fig~\ref{f:spec} suggest that
single Gaussians are not always adequate line models. Hence, we fitted
the spectral features with double Gaussians when necessary.  Each of
the three emission-lines in a given spectrum were fitted separately
and independently. Firstly, we fit the spectra with a 4th order
polynomial avoiding the regions with the emission-lines. We subtract
this continuum contribution and each emission-line we fit with a
single Gaussian. These two fits are used as preparation steps for the
final fit of the emission lines. In the final fit, we fit a double
Gaussian and a 4th order polynomial simultaneously to the original
spectra (not continuum subtracted). The parameters of the continuum
fit and the single Gaussian fit are used as initial values in the
minimisation procedure of the final fit. The results of the continuum
fit were taken as the initial values of the 4th order polynomial
parameters. The initial values of the double Gaussians parameters were
chosen from the single Gaussian fit in the following way: (i) sigma
(width) of the first Gaussian was set to be a half of the sigma of the
single Gaussian fit; the sigma of the second Gaussian was kept the
same as in the fit with the single Gaussian, and (ii) the mean of the
first Gaussian was set to be equal to the mean obtained by the single
Gaussian fit. The mean of the second Gaussian was set to be about 730
km/s (30 \AA) less than the systemic velocity of the galaxy.

To robustly constrain the final fit we introduced the following limits
for the parameters of the first Gaussian: (i) the mean was allowed to
vary 250 km/s around the initial value, and (ii) the sigma was bounded
between 4 and 15 \AA (or 95 km/s and 350 km/s). The lower boundary is
the instrumental resolution. The second Gaussian had the following
constraints: (i) the mean was allowed to vary 500 km/s around the
initial value, and (ii) the sigma was bounded between 4 and 45 \AA (or
95 km/s and 1000 km/s). The upper boundary was seldomly reached. The
final fit with a double Gaussian was performed using the MINPACK
implementation \citep{1980ANL.....80.74M} of the Levenberg-Marquardt
least-square minimisation method\footnote{We used an IDL version of
the code written by Craig B.  Markwardt and available from:
http://astrog.physics.wisc.edu/$\sim$craigm/idl}. The double Gaussian
system increased the goodness of fit, but it was not required at all
positions for all emission-lines. The second Gaussian was needed to
fit the Pa$\beta$ emission-line almost everywhere in the field while
for [FeII] emission, it was necessary only in the centre, and for
[PII] almost not at all.

\begin{figure}
  \includegraphics[width=\columnwidth]{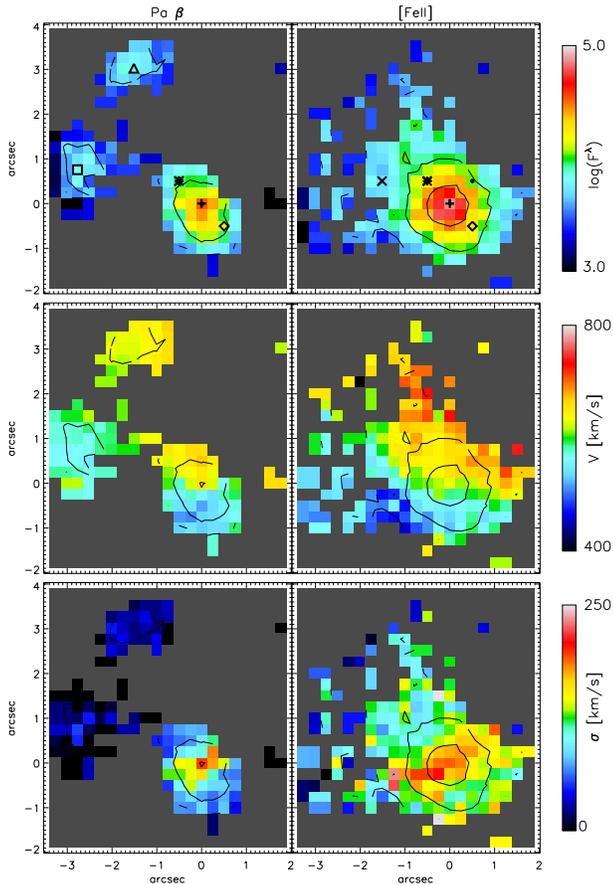}
        \caption{\label{f:maps} Maps of total intensity and kinematics
          of detected ($A/N = 3$) emission-lines. Left: Pa$\beta$;
          right: [FeII]. From top to bottom: the total flux of the
          emission, the mean velocity (centre of the Gaussian), the
          velocity dispersion ($\sigma$ of the Gaussian, without the
          contribution from the instrument dispersion).  Over-plotted
          contours are the contours of the total flux. Different
          symbols relate to spectra in Fig.~\ref{f:lines}. Flux
          $F^\lambda$ is in units of $10^{-17}$
          ergs/s/m$^2$/\AA. Regions without detected emission are shown
          in dark grey.  }
\end{figure}
 
Errors on the parameters of the Gaussians (kinematics) were obtained
through Monte Carlo simulations. For each double Gaussian fit to the
data, we derived uncertainties from 100 random realisations of the
best fit, where the value at each pixel is taken from a Gaussian
distribution with the mean of the best fit and standard deviation
measured on the residuals between the fit and the data. All
realisations together provide a distribution of values from which we
estimate $1\sigma$ (68.3\%) uncertainties.

It is not necessary to expect that binning will always reach the set
levels, since the emission need not to be present, and binning the
noise will not create signal. Calculated amplitudes of both narrow and
broad Gaussians and the standard deviation of the residuals of the fit
can be used {\it a posteriori} to determine the true $A/N$ ratio of
our data. Results show that the reached $A/N$ drops below the targeted
level outside of the central region (size depends on the emission
line). The quickest spatial drop in $A/N$ is seen for [PII], and the
slowest for [FeII] emission-lines.  Also, the highest $A/N$ is
measured for [FeII] and the lowest in [PII] lines.

\citet{2006MNRAS.366.1151S} extensively discuss at which level of
$A/N$ ratio an emission-line is robustly detected. This depends on the
local spectral features (existence of complex absorption features will
require a higher {\it A/N} ratio), but, generally, an emission-line
with $A/N = 2.5 - 3$ can be considered as detected. We binned to
$A/N=2.5$ to keep the spatial resolution as high as possible, but in
{\it a posteriori} check we chose a more conservative limit of
$A/N=3$, and in the rest of the analysis we consider only
emission-lines above this level.

\begin{figure*}
  \includegraphics[width=\textwidth]{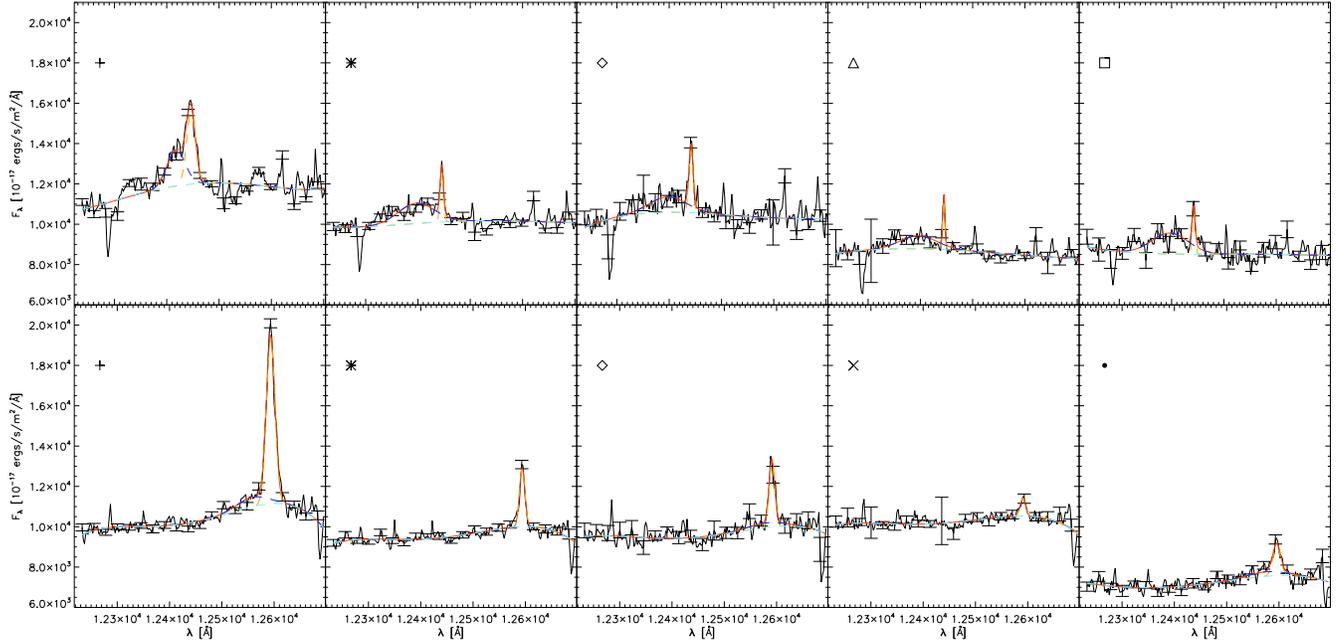}
        \caption{\label{f:lines} Example of spectral features
          recovered from different positions in the nucleus of Cen A
          for different emission-lines. Top: Pa$\beta$; bottom:
          [FeII].  Symbols in top left corner are the same as symbols
          in Fig.~\ref{f:maps}. Every symbol shows the spatial
          location of the bin from which the spectrum was taken.  Only
          regions around the given emission-line used in the fit are
          shown. Yellow, blue and light blue dashed lines are narrow,
          broad Gaussian components and the fit to the continuum,
          respectively. The red line shows the combined fit to the
          data. Note the strong contribution of the broad Gaussian to
          the central Pa$\beta$ emission. Error bars ($1\sigma$) are
          plotted for every 10th data point for clarity.}
\end{figure*}

Emission-lines that met the above criterion are presented in
Figs.~\ref{f:maps} and~\ref{f:lines}. The two rows of
Fig.~\ref{f:maps} show the two-dimensional extent, morphology and
kinematics of Pa$\beta$ and [FeII] emission-lines, respectively from
top to bottom.  The [PII] emission-line reached $A/N = 3 $ level only
in the central 0\farcs5 and the quality of the extracted kinematics
was rather poor, so we do not consider it in the rest of the
study. Fig.~\ref{f:lines} shows examples of the spectral features
obtained from the same spatial positions indicated by the symbols on
both figures, but for different emission-lines (from top to bottom:
Pa$\beta$ and [FeII], respectively).

Although at almost all positions a combination of a narrow and a broad
Gaussian is required to fit the Pa$\beta$ emission-line, only in a few
cases the {\it broad component} can be really recognised as a separate
line.  More often, the {\it broad component} serves as an additional
continuum correction (Fig.~\ref{f:lines}).  When present, the broad
Gaussians are blue shifted with respect to the narrow Gaussians for
(always\footnote{We also fitted the line by fixing the relative
  distance of the Gaussian peaks, which gave equal results with
  marginally worse $\chi^2$.})  $795\kms$. The spatial distribution of
the {\it broad component} (with {\it A/N} $>3$) is limited and
concentrated in the very centre and somewhat elongated in the NE
direction, near the position of the jet and between the blobs of the
narrow component. Line widths are generally less than $800\kms$ (sigma
of the Gaussian), and as such are not comparable to the emission from
classical broad-line regions in AGNs. We call this component {\it
  broad} only to distinguish it from the main line emission and we do
not consider it further.  The main (narrow) component is detected on a
larger area in the centre and at the position of the Pa$\alpha$ blobs,
(Fig.~\ref{f:flux2D}). Spatially, the distribution of the narrow
Gaussians seems to be elongated along the NE direction with $PA
\sim30\degr$.  This value is consistent with the position angle of
\citet{1998ApJ...499L.143S} Pa$\alpha$ disk.

Different results are recovered from the [FeII] line. The broad
Gaussian component to the [FeII] emission-line is almost completely
absent, as expected for a forbidden line, except in the very centre
where {\it A/N} is still below 3 and broad Gaussians improve the fit
to the continuum more than fit a real (robust) emission feature
(Fig.~\ref{f:lines}). The [FeII] line is detectable on a larger area
than Pa$\beta$ and it is not elongated, but more circularly
distributed. The emission is concentrated in the centre, but also
spread towards the NE corner, approximately following the position of
the jet, although it is just above the detection limit and patchy
distributed, mostly over the East Pa$\beta$ blob and approaching the
position of the North blob.

\subsection{Kinematical evidence for a disk}
\label{ss:disk}

Both Pa$\beta$ and [FeII] emission-lines show rotation and they are
likely part of the nuclear disk structure discovered by
\citet{1998ApJ...499L.143S}. Here, we wish to describe their kinematic
properties as they are given on Fig.~\ref{f:maps}.

The there regions with measured Pa$\beta$ emission (the centre, the
East and the North blob), although spatially separated, seem to be
part of the same kinematic structure. The central blob (covering
roughly the central 1 arcsec$^2$) shows a quite regular rotation, with
zero-velocity curve (ZVC) lying horizontally in the East -- West (EW)
direction. The East blob is predominantly blue-shifted, while the
North blob is red-shifted, with respect to the systemic velocity (the
median velocity across the map is $\sim 560\kms$). All three blobs
have similar velocity amplitudes and there is a clear division on the
map: North is receding and the South(-East) is approaching the
observer.  If the emission were detected at all positions, it is
likely that we could see a continuous kinematic map, with a twisting
ZVC: from horizontal (EW direction) to almost $45\degr$ (NE-SW
direction).

Velocity dispersions, measured as the width ($\sigma$) of the narrow
component Gaussian, confirm the relations between the blobs. The East
and North blob have low velocity dispersion as well as the outer
regions of the central blob. The dispersion increases towards the
centre (staying below $200\kms$), possibly being elongated
perpendicular to the orientation of the central blob ($PA \sim
-60\degr$). The low velocity dispersion over most of the region also
suggests a regularly rotating structure, such as a thin disk of gas
clouds.

[FeII] emission covers a larger area and it fills some of the gaps
seen on the Pa$\beta$ maps. The distribution of velocities is
remarkably similar between the two lines. [FeII] emission also shows
rotation (the median velocity across the map is $\sim 590\kms$), which
changes direction: the ZVC is oriented E-W in the centre, but beyond
1\arcsec\, it twists towards NE-SW. Actually, the ZVCs on Pa$\beta$
and [FeII] maps have identical spatial positions and orientations,
where the North on [FeII] map is receding and the South is
approaching. Beyond 2\arcsec where the emission becomes patchy, it is
still possible to trace the rotation of the whole structure.

The velocity dispersion is low in the outer regions, but rises towards
the centre. It is higher than the Pa$\beta$ velocity dispersion
everywhere, except at the edges of the detection region. It also rises
faster (median level is $\sim127\kms$), until levelling at a plateau
($\sim 200\kms$) which is elongated in the same directions as the
enhanced Pa$\beta$ velocity dispersion structure. The orientation of
the plateau ($PA = -60\degr$) is unexpected also because it is not
aligned with the the ZVC. An increase along the ZVC could come from
unresolved rotation in the steepest part of the velocity map (where
velocities change from approaching to receding), but even in this
extreme case the velocity dispersion would not be as elongated. We
will follow this issue further in Sections~\ref{ss:kinkin} and
\ref{ss:res}.

From Fig.~\ref{f:maps} it is obvious that different blobs of Pa$\beta$
and [FeII] emission have related kinematics, even when they are
spatially separated. A similar finding is also visible on SAURON maps
of ionised gas in early-type galaxies \citep{2006MNRAS.366.1151S} and
bulges of Sa galaxies \citep{2006MNRAS.369..529F}, where gas although
patchy or along filaments shows continuous kinematics, suggesting that
all emission elements belong to the same dynamical structure, but that
the level of ionisation, the amount of gas or extinction is not the
same and varies locally. 

The rotation of both lines is to a first order regular and suggests
that the emission is distributed in a disk-like structure. It is also
clear that the kinematics of Pa$\beta$ and [FeII] emission are
generally similar (velocity orientation and amplitude), but there are
some notable differences such as a higher velocity dispersion in
[FeII] lines.  Judging from velocity dispersion maps, it seems that
[FeII] emission is more perturbed than Pa$\beta$ and its motion is not
completely ordered nor purely disk-like. Deviations could be caused by
physical processes such as inflows, outflows, non-axisymmetric
perturbations in the potential, warps or simply material not belonging
to the disk. We investigate this in more details below.

\begin{figure*}
    \includegraphics[width=\textwidth]{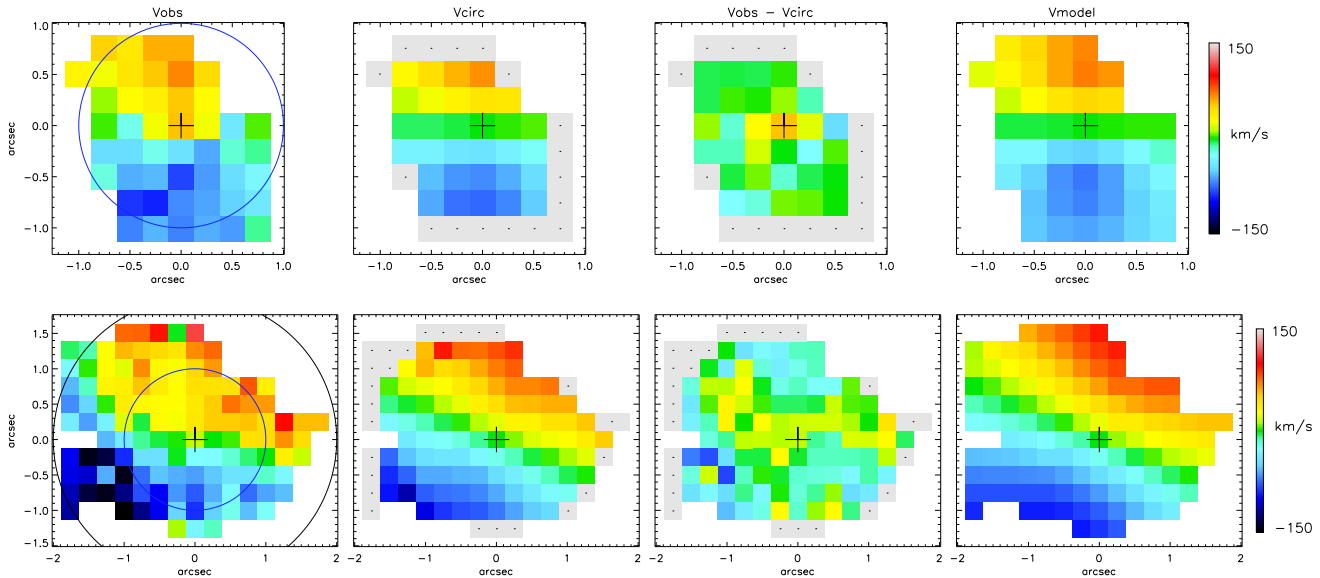}
    \caption{\label{f:model} Maps of the observed mean velocity (far
      left), circular velocity (middle left), residuals between the
      mean and circular velocity (middle right), and the best fitting
      thin disk models (far right) using the derived orientation for
      Pa$\beta$ (top) and [FeII] (bottom) emission-lines disks.
      Circles over-plotted on observed velocity maps have radii of
      0\farcs9 (blue) and 1\farcs8 (black). Bins marked with black
      dots were not included when constructing circular velocity maps.
      The plus symbol denotes the origin of the map, assumed to be the
      position of the Cen A nucleus.  Standard deviations of residual
      maps are $18\kms$ and $25\kms$ (but $16\kms$ within the blue
      circle), for Pa$\beta$ and [FeII] maps, respectively. The method
      to construct circular velocity maps is described in
      Section~\ref{ss:kinkin} and the thin disk model is described in
      Section~\ref{s:dyn}. }
\end{figure*}

\subsection{Orientation of the velocity structures}
\label{ss:kinkin}

Assuming that the velocities represent material moving in a thin disk,
it is, in principle, possible to determine both $PA$ and inclination
($i$) at the same time and as a function of radius.  This is, in
practice, often quite difficult, especially if there are deviations
from circular rotation or data are noisy. In our case, we decided to
determine only global properties and to do that through a two stage
process.

We first follow the method for determination of the global kinematic
$PA$ from Appendix C of \citet{2006MNRAS.366..787K}.  Briefly, one
constructs a set of bi-anti-symmetric velocity maps (because of the
assumed axisymmetry) with different $PAs$ and compare them with the
observed velocity map. The likeness of a model map ($V_{m}$) to the
observed map ($V$) is measured by the quantity $\chi^2 = \sum^{N}_j
[(V_{m,j} -V_j)/\Delta V_j]^2$, where $\Delta V_j$ are measurement
errors on velocities for each bin $j$ on a map $m$. The model map with
the $PA$ closest to the observed value will contribute with the smallest
$\chi^2$.

We applied this method to [FeII] and Pa$\beta$ maps which were
slightly modified by selecting only spatially continuous emission on
the maps. This meant decreasing the size of maps to the central
regions (about 1 arcsec$^2$ for Pa$\beta$ and about 2 arcsec$^2$ for
[FeII]), but increased the robustness of the method. For these trimmed
maps (see the first column in Fig.~\ref{f:model}) we obtained the
following values: $PA = - 3\degr \pm 10\degr$ for Pa$\beta$ and $PA =
- 23\degr \pm 3\degr$ for [FeII] velocity maps.  All errors quoted
here are at $3\sigma$ confidence level. The [FeII] map is about 2
times larger than the Pa$\beta$ map and difference in $PA$ comes from
the change in the ZVC as one includes a larger area. On
Fig.~\ref{f:model} the blue and black circles on the maps show the
regions used in the fit to determine the $PA$ for Pa$\beta$ and
[FeII], respectively. Repeating the exercise for [FeII] within the
blue circle, one gets $PA = -14\degr \pm 5\degr$, which is in good
agreement with the Pa$\beta$ value.

The next step is to estimate the inclination of the disk assuming that
the motion in the disk is purely circular. In that case, a velocity
profile extracted along an ellipse will be described by a cosine law,
if the ellipse samples equal radii in the disk plane and has the same
$PA$ as the disk. This ellipse is then related to the inclination
through its ellipticity: $1-\epsilon = \cos i$.  Using the {\it
  kinemetry} software of \citet{2006MNRAS.366..787K} and assuming a
constant $PA$ measured above, we estimate the $i$ of the maps in the
following way. Starting from a given $i$ we stepped through the
velocity map extracting velocity profiles along ellipses with
$1-\epsilon = \cos i$ and increasing semi-major axis $R$. Each
velocity profile we least-squares fitted with the formula $V(R,
\theta) = (V_c(R) / \sin i) \cos \theta$, where $V_c$ is the circular
velocity and $\theta$ eccentric anomaly, measured from the projected
major axis of the ellipse (given by the fixed $PA$). We linearly
interpolated $V(R,\theta)$ to $V(x,y)$ creating a model velocity map
at the observed coordinates on the sky. This has been done for a range
of different inclinations and each model velocity map ($V_m$) was
compared to the observed map ($V$) similarly as above using a
quantity: $\chi^2 = \sum^{N}_j [(V_{m,j} -V_j)/\Delta V_j]^2$, where
$\Delta V$ are measurement errors. The inclination of the model map
with the minimum $\chi^2$ was recognised as the global estimate of the
inclination of the observed disk. An almost identical procedure (they
varied both $PA$ and $i$ simultaneously) was used previously by
\citet{2006A&A...447..441B}.

In the case of Pa$\beta$ emission, setting the $PA$ to $-3\degr$ we
obtained $i=25\degr \pm 5\degr$.  Setting the $PA$ to $-23\degr$ for
[FeII] disk we obtained $i=30\degr \pm 5\degr$. The two inclinations
are consistent with each other, and are also consistent with studies
(\cite{2001ApJ...549..915M}, \cite{2005AJ....130..406S},
\cite{2006A&A...448..921M}), but somewhat smaller than models by
\citep{2006ApJ...643..226H}.  In the second column of
Fig.~\ref{f:model} we show the reconstructed circular velocity maps
using the best fitting $PA$ and $i$.

\begin{figure}

  \includegraphics[width=\columnwidth]{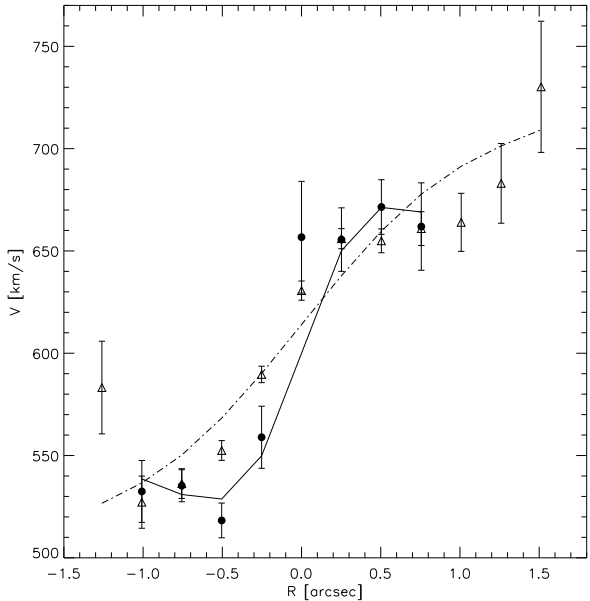}
     \caption{\label{f:slits} Measured emission-line velocities and
       model predictions (see Section~\ref{s:dyn} for details) along
       $PA=0\degr$. The [FeII] velocity data and the best fitting
       model are shown with triangles and the dashed-dotted line,
       respectively.  The Pa$\beta$ velocity data and the best fitting
       model are shown with filled circles and the black line.}
\end{figure}

One can most clearly see the deviations from a pure circular motion on
the residual maps obtained by subtracting the reconstructed circular
velocity maps from the observed velocities (shown in column 3 on
Fig.~\ref{f:model}). Both residual maps have visible non-zero
components. On average, residuals are higher on the [FeII] map
(standard deviation is $18\kms$ for Pa$\beta$ and $25\kms$ for
[FeII]), but within the central region (blue circle) the residual are
very similar and have the same spatial trend )standard deviation of
$\sim 16 \kms$). This is consistent with the rising [FeII]/[PII] ratio
towards the edge of the map, where shocks become an increasingly
important factor in the ionisation mechanism, and possibly influence
the kinematics. Shocks can result from the jet interaction with the
surrounding matter, but can also occur between the tilted rings of the
warped disk. On larger scales ([FeII] residuals map) there is a hint
of a more complicated pattern which primarily arises from the
difference between the observed (changing) and assumed (constant)
orientation.  This, possibly three-fold, pattern can be associated
with inflows, outflows or warps.

The most distinct feature on the velocity maps is the twist of the ZVC
beyond 1\arcsec. Inside this radii the black hole is dominating the
potential \citep{2006A&A...448..921M} and is able to stabilise gas
clouds in a disk. The central blob of Pa$\beta$ emission is largely
confined to this region only, but iron is ionised in almost continuous
gas clouds up to 2\arcsec, and the influence of instabilities is
visible in the central emission-line disk.  The two other Pa$\beta$
blobs at $~\sim 2.5\arcsec - 3\arcsec$ from the nuclei, behave
similarly like [FeII] disk on large radii, being far from the sphere
of influence of the black hole where the instability can perturb the
disk.

The dusty disk in Cen A is known to be warped at kpc scales
\citep{1987MNRAS.228..595B,1992ApJ...391..121Q,1992ApJ...387..503N,
  1993ApJ...412..550Q}. There are evidences for a bar spiral system
\citep{1999A&A...341..667M, 2000A&A...353...72B, 2002ApJ...565..131L},
although the recent infrared observations support the warp disk
hypothesis \citep{QU06}. Our data shows that the warp (or a bar-like
perturbation) can be traced all the way to 20 pc scales.  It is,
however, difficult to be conclusive on the matter from these data, and
the meaning of residuals has to be taken with care, especially when
the measured uncertainties are taken into account: $16\kms$ for
Pa$\beta$ and $19\kms$ for [FeII]. We will, however, following the
recent results, continue in the framework of the warped disk.

Comparing the orientation and the extent of the velocity dispersion
plateau with the tilted ring model of \citet{1992ApJ...387..503N} and
\citet{QU06}, we suggest that the origin of the high velocity
dispersion is in the unresolved rotation coming from different folds
of the warped disk. We draw this conclusion from Figs.~10 and 11 of
\citet{1992ApJ...387..503N} and Figs.~5 and 7 of \citet{QU06}, which
show that the disk changes its orientation such that in the inner part
of the disk ($r < 60\arcsec$) its SW points and in the outer part ($r
> 100\arcsec$) its NE points are closest to the observer. Although
both studies construct the warped disk model to larger radii (beyond
40\arcsec) than what we present in this work, the orientation of the
high velocity dispersion plateau ($PA = -60\degr$ or $PA = 120\degr$)
agrees well with the measured orientation of the warped disk. This
means that our line-of-sight crosses the disk a few (at least two)
times, and the observed kinematics will depend on where the ionisation
of the gas clouds occurs.

The [Fe II] emission-line arises from partially ionised regions, i.e.
regions with a few per cent fraction of H{\small I} in the H{\small
  II} state. These regions are typically seen in the narrow line
region environment of an AGN, where X-ray photons from the central
engine can penetrate deep into the cloud. Thus, partially ionised
regions do not have a direct view of the central engine, being
shielded by a column of~$\sim 10^{20}$ cm$^{-2}$ of $N_H$. The
observed [FeII] emission could originate in different folds of the
warped disk, also further away from the central engine. The resulting
large difference in velocities along the line-of-sight, observed as
unresolved rotation, has two effects: the velocity gradient is
shallower and the velocity dispersion higher than in the case of
resolved rotation.

In Fig.~\ref{f:slits} we compare the central velocity gradient between
the Pa$\beta$ and [FeII] emission-lines, extracted from velocity maps
along a slit of 3\arcsec\, length, 0\farcs25 width and
$PA=0\degr$. Within $0\farcs5$, Pa$\beta$ emission shows a steeper
rotation curve than [FeII] emission, while around 1\arcsec\, both
emission-lines reach the same level. In the substantially warped gas
disk, the innermost regions would provide the required shielding for
the [FeII] emitting material slightly further out.  The lack of [FeII]
emission from the innermost region, in contrast to Pa$\beta$, would
explain a slower rise of the [Fe II] line-of-sight velocity curve with
increasing radius, plotted in Fig.~\ref{f:slits}. Also, unlike
Pa$\beta$, the [FeII] emitting material is confined to specific
regions of the gas disk, increasing the likelihood that it provides an
incomplete picture of the gas kinematics in the central region.  The
higher velocity dispersion indicate a more turbulent medium, one in
which shocks may be playing an important role.  This too, raises
doubts about the [FeII] emitting gas serving as a good tracer of the
gravitational potential in the innermost nuclear regions.

%
%

\section{Evidence for the central black hole}
\label{s:dyn}

In the previous section we showed evidence that the observed ionised
gas is moving in a disk-like structure, which is consistent with being
warped beyond 1\arcsec. In this region, the motion is not completely
ordered, but the level of the residuals is comparable with measurement
uncertainties. We also showed that [FeII] emission-line kinematics
suffers from observed unresolved rotation, which can be explained with
warped disk morphology. We continue the analysis by constructing
simple dynamical models, based on the observed properties of the disk
(orientation, circular motion) with a purpose of testing the
hypothesis that there is a central black hole in Cen A that can
explain the observed kinematics.

\subsection{Mass models}
\label{ss:mass}

As reported by \citet{2005AJ....130..406S}, the enclosed stellar mass
within the central 1\arcsec\, is $3\times 10^7 M_{\odot}$, for a
deredened profile, and $2\times 10^7 M_{\odot}$ for an observed (not
deredened) light profile. The mass of the black hole expected from the
M$_\bullet - \sigma$ relation \citep[as given
  by][]{2002ApJ...574..740T} is $3\times10^7 M_{\odot}$. The mass of
the central black hole in Cen A determined by previous studies is,
however, around $1\times10^8 M_{\odot}$. Assuming a velocity
dispersion of $\sim150\kms$ \citep{2005AJ....130..406S}, the sphere of
influence of this object is $\sim 1\farcs3$. This suggest that within
1\arcsec\, the dominant contribution to the potential comes mostly
from the central black hole. The total potential can then be written
in the following way:
\begin{equation}
\label{eq:pot}
\Phi = \Phi_\star + \Phi_{\bullet},
\end{equation}

\noindent where the black hole contribution is simply given by
$\Phi_{\bullet} = G M_{\bullet}/r$, while $\Phi_\star$ can be derived
from the stellar luminosity density assuming a mass-to-light ratio $
M/L$. The stellar luminosity density of a system is obtained by
deprojecting the observed surface brightness distribution, assuming a
symmetry type (e.g.  axisymmetry) for the system and its
inclination. This can be conveniently done using the the
Multi-Gaussian expansion method
\citep[MGE;][]{1992A&A...253..366M,1994A&A...285..723E}.

\begin{figure}
  \includegraphics[width=\columnwidth]{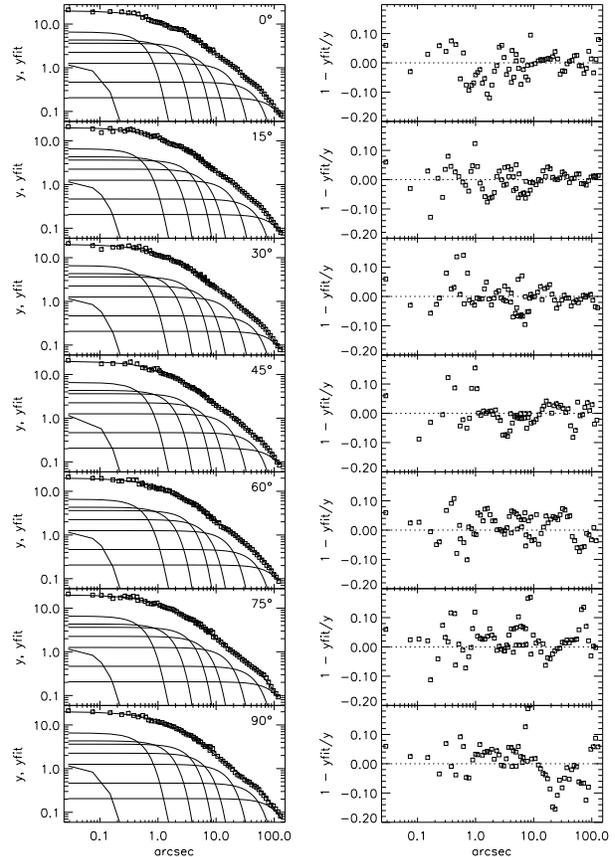}
     \caption{\label{f:mgefit} 
       Comparison between the combined NICMOS/F222M and 2MASS K-band
       photometry (open squares) with the MGE model (line) in seven
       angular sections as a function of radius. The right panels
       show the radial variation of the relative error along the
       profiles. }
\end{figure}

Using the MGE software of \citet{2002MNRAS.333..400C}, we
simultaneously fitted a ground-based K-band image from 2MASS Large
Galaxy Atlas \citep{2003AJ....125..525J} ( 3\arcsec $<$ r $<$
200\arcsec) and the space-based NICMOS F222M image
\citep{1998ApJ...499L.143S} (0\arcsec $<$ r $<$ 9\arcsec). The 2MASS
image was scaled to the NICMOS image, while the sky was measured on
the 2MASS image. The MGE software analytically deconvolves the
instrumental PSF from the observed light, which we determined using
the TinyTim software \citet{TinyTim} and also parameterised by three
two-dimensional Gaussians, normalised such that their amplitudes,
$G_j$, add up to unity. Following the suggestion in
\citet{2002MNRAS.333..400C}, we increased the minimal axial ratio of
the Gaussians, $q_j$, while not changing $\chi^2$ significantly, in
order to make as large as possible the range of permitted
inclinations.  Parameters describing the parametrisation of the PSF
and the deconvolved MGE model are presented in Tables~\ref{t:psf} and
\ref{t:mge}, while a comparison of the combined ground and space-based
light profiles and the MGE model is shown in Fig.~\ref{f:mgefit}.

\begin{table}
   \caption{The MGE parameters of the circular PSF.}
   \label{t:psf}
$$
   \begin{array}{ccc}
       \hline
       \noalign{\smallskip}
       $j$ & $G$_{j}  & \sigma_{j} $(arcsec)$ \\
       \noalign{\smallskip}
       \hline
            1 & 0.754 & 0.041\\
            2 & 0.166 & 0.126\\
            3 & 0.080 & 0.508\\
       \noalign{\smallskip}
       \hline
   \end{array}
$$
\end{table}

\begin{table}
    \caption{The parameters of the MGE model of the deconvolved K-band
   surface brightness of Cen A. 
  }
  \label{t:mge}
$$
   \begin{array}{ccccc}
       \hline
       \noalign{\smallskip}
       $j$ & $I$_{j} (L_{\odot}{pc}^{-2}) &\sigma_{j} $(arcsec)$ &$q$_{j} & $L$_{j} (\times10^{9} L_{\odot}) \\
       \noalign{\smallskip}
       \hline
 1 & 70122.1720  &   0.0506     & 0.9500 &   0.00030887\\
 2 & 113639.0196 &    0.4428    &  1.0000&   0.040313\\
 3 &  68662.2409 &    1.2923    &  1.0000&   0.20744\\
 4 &  56428.6385 &    2.9233    &  1.0000&   0.87241\\
 5 &  34809.6545 &    5.3776    &  1.0000&   1.8211\\
 6 &  19386.9662 &   13.6883    &  0.9500&   6.2431\\
 7 &   7184.3315 &   37.4720    &  1.0000&   18.250\\
 8 &   3174.4729 &   92.0318    &  0.9823&   47.783\\
       \noalign{\smallskip}
       \hline
   \end{array}
$$
\end{table}

\subsection{Simple dynamical models}
\label{ss:hole}

The basic assumption of our simple dynamical models is that clouds of
ionised gas move in circular orbits within a thin disk. This is an
often used approach pioneered for black hole studies by
\citet{1997ApJ...489..579M} and subsequently applied to the ionised
gas disk in Cen A by other authors \citep{2001ApJ...549..915M,
  2006A&A...448..921M, 2006ApJ...643..226H}.

The circular velocity is given by $V_c (R) = \sqrt{R (\ud \Phi/\ud
  R)}$, where the gravitational potential $\Phi$ is as in
  eq.~(\ref{eq:pot}). If mass-to-light ratio {\it M/L} is taken to be
  zero, the gravitational potential is described only by a
  'point-mass' contribution ($\Phi_\bullet$), and the circular
  velocity simplifies to $V_c (R) = \sqrt{G M_\bullet/R} $. This model
  we call {\it Kepler}, while the more complex model, which includes
  the contribution of the stellar potential we call {\it MGE}. The
  calculation of the potential from the MGE parametrisation in that
  case is done as described in \citet{2002ApJ...578..787C}.

Thin disk models assume that the motion is purely circular and that
the velocity dispersion of the gas is negligible. Our observations of
the Pa$\beta$ disk are consistent with this assumption, but it breaks
in the case of the [FeII] disk. We speculate that the increase in the
velocity dispersion is related to the location of the [FeII] emission
in the disk, the geometry of the warped disk and the turbulent medium
with increasingly important [FeII] ionising shocks waves (see
Section~\ref{ss:res} for details). From this reasons, our observations
of [FeII] emission are not good tracers of the central potential and
are not suitable for the determination of the black hole mass with the
presented dynamical models. In what follows, we focus only on the
Pa$\beta$ emission, but we also construct a model for the [FeII] disk
for comparison with observations (Figs~\ref{f:model}
and~\ref{f:slits}).

To accurately reproduce the observed kinematics we include in the
model the parametrisation of the emission-line surface brightness
obtained by fitting a function to the measured flux, taking the $PA$ and
the $i$ of the gas disk into account. We tested different
functional forms, finally choosing the following one which describes
the spatial distribution of the emission-lines surface brightness
well:
\begin{equation}
\label{eq:emsb}
I = I_{0} + I_{1} \exp{-\frac{r}{r_{1}} },
\end{equation}

\noindent where $r_1$ is the scale radius and $I_0$ and $I_1$ are the
scale factors for the given gas component. The best fitting values for
the Pa$\beta$ assuming for the orientation of the disk ($PA=-3\degr,
i=25\degr$) are $I_0=1844.4$, $I_1=32884.9$ and $r_1 =
0\farcs503$. For other orientations we obtain slightly different
values and use those in the corresponding dynamical models.

When constructing dynamical models, we fix the geometry of the disk,
use the above mentioned emission-line surface brightness
parametrisation, and vary parameters defining the gravitational
potential: {\it M/L} ratio and $M_{\bullet}$. The resulting models of
the circular velocity maps, convolved with the PSF and the pixel size
of the CIRAPSS observations \citep{1995MNRAS.274..602Q}, are compared
with the observed models.  The best fitting model is determined by
minimising the $\chi^2 = \sum^{N}_j [(V_{m,j} -V_j)/\Delta V_j]^2$,
similarly as done before.

\begin{table}
   \caption{The parameters of the best fit thin disk models}
   \label{t:model}
$$
   \begin{array}{ccccccc}
       \hline
       \noalign{\smallskip}
       $model$ & $mass$ & $i$  &$PA $ & M_{BH} & $M/L$ & \chi^2\\
               &$model$&$[degr]$&$[degr]$&$[$ 10^7  $M$_{\bullet}$]$& $[K band]$ &\\
       \noalign{\smallskip}
       \hline
       $Pa$\beta\\
       \hline
       \noalign{\smallskip}
        p1 & $MGE$    & 25 & -3  & 8.25^{+2.25}_{-4.25} & 0.0 + 1.3          & 33.5\\
        p2 & $MGE$    & 30 & -23 & 6.5^{+3.5}_{-2.5} & 0.45^{+1.0}_{-0.45}   & 77.0\\
        p3 & $MGE$    & 25 & -15 & 9.0^{+3.5}_{-3.0} & 0.15^{+1.55}_{-0.15}  & 46.9\\
        p4 & $MGE$    & 45 & -27 & 3.5^{+2.0}_{-2.5} & 0.35^{+0.85}_{-0.35}  & 104.3\\
      
        p5 & $Kepler$ & 25 & -3  & 8.25^{+2.00}_{-1.75}  & 0.0           & 33.5\\
        p6 & $Kepler$ & 30 & -23 & 8.0^{+2.0}_{-2.0}     & 0.0           & 78.1\\
        p7 & $Kepler$ & 25 & -15 & 9.5^{+2.0}_{-1.0}     & 0.0           & 47.0\\
        p8 & $Kepler$ & 45 & -27 & 5.0^{+1.3}_{-1.0}     & 0.0           & 107.9\\
       \noalign{\smallskip}
       \hline

   \end{array}
$$ {Notes -- We used two mass models: {\it Kepler} where the potential
     is given by the central point-mass, a mass model obtained from
     our {\it MGE} parametrisation of the stellar light. There were 47
     velocity data points used to constrain the models.}
\end{table}

\begin{figure}

  \includegraphics[width=\columnwidth]{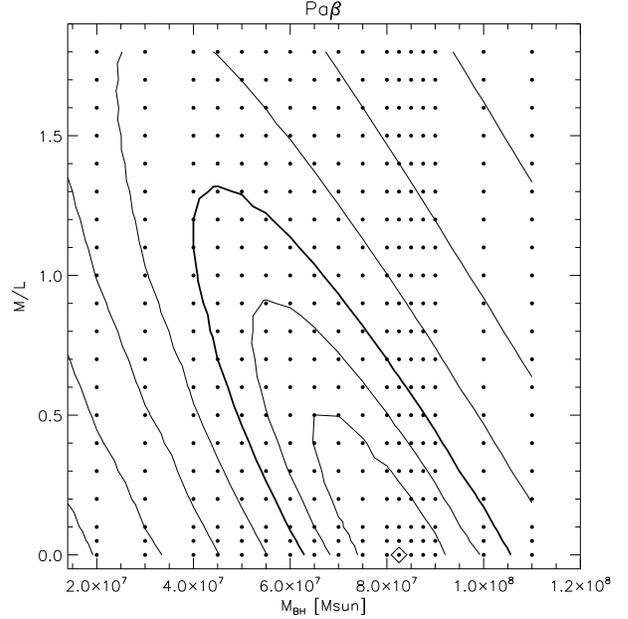}
     \caption{\label{f:chi2grid} A grid of models with different mass
       of the black hole ($M_{BH}$) and mass-to-light ratio ($M/L$)
       for the Pa$\beta$ disk at $i=25\degr$ and
       $PA=-3\degr$. Over-plotted are the contours of the constant
       $\chi^2$. The first three contours correspond to 68.3\%, 95.4\%
       and 99.73\% (thick contour) confidence regions for a
       distribution of 2 degrees of freedom.  The best fitting model
       is shown with a diamond.}
\end{figure}

\subsection{Results and discussion}
\label{ss:res}

We calculated a range of dynamical models with different disk
orientations, black hole masses and mass-to-light ratios. We applied
the disk orientation from Section~\ref{ss:kinkin} for Pa$\beta$
emission-line disk and also used values from the literature
\citep{2006A&A...448..921M, 2006ApJ...643..226H}. 

Table~\ref{t:model} presents the values of the best fitting models.
Our best fitting orientation of the Pa$\beta$ disk yields a model with
$M_{\bullet}= (8.25^{+2.25}_{-4.25}) \times 10^7 M_{\odot}$ and $M/L=0.0 \pm
1.3$. Uncertainties quoted for all presented models are at the
$3\sigma$ level. The expected value of $M_\bullet$ from the $M_\bullet
- \sigma$ relation is excluded at a $3\sigma$ confidence level, taking
into account the scatter of the relation of 0.34 dex
\citep{2005SSRv..116..523F}, but our estimate, within its large
uncertainties, is consistent with other measurements in the literature
at similar inclinations \citep{2006A&A...448..921M,
2006ApJ...643..226H}. Values for $M_{\bullet}$ obtained from stellar
kinematics \citep{2005AJ....130..406S} are still higher (even their
model for $i=20\degr$) and formally excluded at a $3\sigma$ level. The
estimated $M/L$ ratio of the best fitting model is consistent with
being zero confirming that within 1\arcsec\, the gravitational
potential is almost entirely determined by the black hole and not
dependant much on the parametrisation of the stellar light.

Confidence levels over-plotted on the grid of different models with
our best fitting disk orientation are presented in
Fig.~\ref{f:chi2grid}. The reconstructed velocity map of the best
model is shown in Fig.~\ref{f:model} and a cut along $PA=0\degr$ in
Fig.~\ref{f:slits}. For completeness, we report here that using our
best fitting orientation for the [FeII] disk we get $M_{BH}=
(0.1^{+1.0}_{-0.1}) \times10^7 M_{\bullet}$ and $M/L=2.0 \pm 0.4$. The
untrustworthiness of this model can be seen in Figs.~\ref{f:model}
(last panel) and~\ref{f:slits} (dash-dotted line).

In the construction of the thin disk models the orientation of the
disk is crucial, and the final parameters of the potential
($M_{\bullet}$, $M/L$) will depend on it. We constructed other
dynamical models (Table~\ref{t:model}) using a set of disk
geometries. For Pa$\beta$: the best orientation obtained from [FeII]
kinematics (model p2), the best disk orientation from
\citet{2006A&A...448..921M} (model p3) and geometry of the best model
from \citet{2006ApJ...643..226H} (model p4). Model p5 has the best
Pa$\beta$ disk geometry obtained from the kinematics, while models p6
-- p8 have the same geometries as above (in the same order), but these
mass models include only the contribution from the black hole.

In the last column of Table~\ref{t:model}, $\chi^2$ values for each
model are listed. The best model is p1, with a geometry determined
from the kinematics derived in Section~\ref{ss:kinkin}. Models p1 and
p5 are formally equally good, but other models (including p3 and p7)
are excluded at the $3\sigma$ level (for 2D distribution). The models
are more sensitive to the inclination than to a change in $PA$. Models
with Keplerian and MGE mass parametrisation give similar results, but
the uncertainties in the estimated mass are somewhat smaller in
Keplerian models, which have a fixed M/L.

As an additional check of our results, we also constructed models with
a given $M_\bullet$ and $M/L$ with the purpose to recover the measured
disk orientations( $PA$, $i$). We chose models with parameters from
the literature: \citet{2006ApJ...643..226H} -- model pg1,
\citet{2006A&A...448..921M} -- model pg2, \citet{2005AJ....130..406S}
-- model pg3. The models pg4 is a hybrid combining the value from the
M$_\bullet - \sigma$ relation and the $M/L$ ratio from other
studies. Results are given in Table~\ref{t:model_geo}.

\begin{table}
   \caption{Additional disk models for given M$_\bullet$ and $M/L$ ratio.}
   \label{t:model_geo}
$$
   \begin{array}{ccccccc}
       \hline
       \noalign{\smallskip}
       $model$ & $mass$ & M_{BH} & $M/L$ &  $i$  &$PA $ & \chi^2\\
               &$model$ &$[$ 10^7  $M$_{\bullet}$]$& $[K band]$ &$[degr]$&$[degr]$&\\
       \noalign{\smallskip}
       \hline
       $Pa$\beta\\
       \hline
       \noalign{\smallskip}
        pg1 & $MGE$ &   6.1 & 0.7  & 25 \pm 5     & -5^{+10}_{-10}      & 37.3\\
        pg2 & $MGE$ &  11.0 & 0.0  & 20 \pm 5     & 0^{+9}_{-13}      & 34.6\\
        pg3 & $MGE$ &  24.0 & 0.68 & 20 \pm 1     & -20^{+7}_{-6}     & 128.6\\
        pg4 & $MGE$ &   3.0 & 0.7  & 30^{+10}_{-7}& 0^{+7}_{-15}  & 44.4\\
      \noalign{\smallskip}
      \hline

   \end{array}
$$ 
\end{table}

As with previous models, most configurations give similar goodness of
fit values, only model pg3 has a very different $\chi^2$ and can be
discarded.  All models recover mutually consistent (within errors)
geometrical parameters which are also in good agreement with values
obtained in Section~\ref{ss:kinkin}. Model pg2
(\cite{2006A&A...448..921M} geometry) is expectedly the closest to our
best fitting model (p1) in terms of $\chi^2$. Assuming M$_\bullet$
from the \citet{2002ApJ...574..740T} relation gives a model that is
close to be discarded with $3\sigma$ confidence level.

%
%

\section{Conclusions}
\label{s:con}

This paper presents observations of the active elliptical galaxy,
NGC~5128 (Cen A), obtained with the integral-field spectrograph
\cirpass. In the spectra there are three emission-lines, [PII], [FeII]
and Pa$\beta$, with different spatial distributions and relative
strengths.  All lines are present in the central 1\arcsec, [FeII]
being the most extended. Pa$\beta$ is also detected in two blobs,
North and East of the nucleus. These blobs are in an excellent
agreement with Pa$\alpha$ blobs discovered by the HST
imaging. Analysis and results are summarised in the following items.

\begin{itemize}
  
\item The reconstructed total intensity image shows two regions with
  strong continuum contribution: in the centre and NE of the
  centre. The NE region overlaps with the position of the radio jet
  (in projection) and is located between the Pa$\beta$ blobs. The
  region can also be associated with the knot N1 of the radio
  jet. This suggests a possible jet-induced star forming region in Cen
  A on scales of $\sim30$ pc, in addition to the well known large
  scale (1kpc) jet-induced star forming region.

\item The location of the East Pa$\beta$ blob coincides with the kink
  in the radio jet suggesting a direct link between the two
  phenomena. If one assumes that the North Pa$\beta$ blob was also
  induced by the jet, the position of the blobs show that the jet
  precessed for about 25\degr in projection.
 
\item We detect [PII] emission, which can be used together with [FeII]
  emission to discriminate between photo- and shock-induced ionisation
  of gas clouds. The [FeII]/[PII] ratio rises steadily with radius,
  being close, but above pure photo-ionisation levels. This suggests
  that shocks are more important with increasing distance from the
  centre.
  
\item We determined spatially resolved kinematics from Pa$\beta$ and
  [FeII] emission-lines. Although the emission (of both lines) is
  divided in blobs or it is patchy at certain regions, it has a well
  defined and uniform sense of rotation and spatially separated
  emission regions are part of the same kinematic structure. The
  velocity dispersion is low for Pa$\beta$, but high for [FeII]
  emission with a plateau oriented at $\sim -60\degr$. The low
  dispersion is suggestive of a motion in a regular disk for
  Pa$\beta$, but the plateau in the [FeII] velocity dispersion is a
  clear evidence of a more complex distribution of ionised iron.
  
\item Kinematics of both emission-lines is remarkably similar in
  appearance, as traced by ZVC: central regions are rotating around an
  EW axis, while at larger radii there is a twist in ZVC and the
  rotation is around NE -- SW axis. The twist in ZVC suggest
  contribution from a bar-like perturbation or a warp in the disk on
  20 pc scales.
  
\item We test the assumption that the emission in the central
  1-2\arcsec\, is confined to a disk by constructing maps of circular
  velocity and comparing them with observations. This process is
  divided in two steps. We first determine the $PA$ and (using this
  position angle) the inclination of the assumed disk. The best
  fitting parameters are: $PA = -3\degr \pm 10\degr$ and $i=25\degr
  \pm 5$ for Pa$\beta$ and $PA = -23\degr \pm 3\degr$ and $i=30\degr
  \pm 5$ for [FeII]. Using these values we extract circular velocity
  component and remove it from the data.  The residual map of
  Pa$\beta$ velocities is noisy without recognisable pattern and
  residuals are largely within measurement errors. [FeII] residuals
  are more complicated, showing a three-fold pattern and being
  somewhat larger than measurement uncertainties. This can be traced
  to the twist in the ZVC. We conclude that the emission in the
  central 1\arcsec, the region which is likely to be within the sphere
  of influence of the central black hole, is consistent with being in
  circular motion in a disk, but outer regions show departures which
  could be attributed to perturbations (such as a bar) or to a warp
  extending inwards from 1 kpc scales (as previously detected in Cen
  A) to 20 pc scales.

\item A shallower rotation curve of the [FeII] emission is likely
  related to the high velocity dispersion and the geometry of the
  warped disk. We speculate that the excitation of [FeII] line happens
  also at larger radii from the central engine and the line-of-sight
  passes two times through the warped disk, such that at the spatial
  resolution of our observations the data contain a contributions of
  gas moving in two different directions (lowering the velocity, but
  increasing the velocity dispersion). This is supported by 
  coinciding orientations of the high velocity dispersion region and
  the warped disk. Following this, we do not consider our [FeII]
  kinematics for determining the mass of the central black hole.

\item We construct simple dynamical models of the Pa$\beta$
  emission-line disk to determine the mass of the central black
  hole. For the geometry of disks we use kinematically determined
  orientation parameters. We measure M$_\bullet = 8.25^{+2.25}_{-4.25}
  \times 10^{7}$ M$_{\odot}$. This value is consistent with previous
  measurements, while the predicted value of the black hole mass from
  the M$_\bullet - \sigma$ relation is excluded at the $3\sigma$ level
  (for a 2D distribution), taking into account the scatter of the
  relation.
  
\item We construct models with different geometrical orientations
  taken from previous studies, which give similar results, being
  marginally less good, but often statistically indistinguishable. We
  also construct models using literature values for M$_\bullet$ and
  the value predicted by the M$_\bullet-\sigma$ relation, but leaving
  geometrical parameters free.  These models confirm our original
  results.

\end{itemize}

\vspace{+1cm}
\noindent{\bf Acknowledgements}\\

We thank M. Hardcastle for kindly providing radio and X-ray images of
Cen A and Marc Sarzi and Roger Davies for helpful discussions. RGS is
greatfull to Keith Shortrage for providing the conceptual background
for \emph{SCRUNCHING} spectra by locating the code for the FIGARO
REBIN task. NT is supported by a Marie Curie Excellence grant from the
European Commission MEXT-CT-2003-002792 (SWIFT). This paper is
partially based on observations obtained at the Gemini Observatory,
which is operated by the Association of Universities for Research in
Astronomy, Inc. (AURA), under a cooperative agreement with the
U.S. National Science Foundation (NSF) on behalf of the Gemini
partnership: the Particle Physics and Astronomy Research Council
(PPARC, UK), the NSF (USA), the National Research Council (Canada),
CONICYT (Chile), the Australian Research Council (Australia), CNPq
(Brazil) and CONICET (Argentina).  This publication makes use of data
products from the Two Micron All Sky Survey, which is a joint project
of the University of Massachusetts and the Infrared Processing and
Analysis Center/California Institute of Technology, funded by the
National Aeronautics and Space Administration and the National Science
Foundation. Based on observations made with the NASA/ESA Hubble Space
Telescope, obtained from the data archive at the Space Telescope
Institute. STScI is operated by the association of Universities for
Research in Astronomy, Inc. under the NASA contract NAS 5-26555. We
thank the Raymond and Beverly Sackler Foundation and PPARC for funding
the CIRPASS project.\\


\label{lastpage}
\end{document}